\documentclass[pra,twocolumn,floatfix,superscriptaddress,nofootinbib]{revtex4-1}

\usepackage{graphicx,amsbsy,amsmath,bm,tikz}
\usepackage[version=4]{mhchem}
\usepackage{orcidlink}
\usepackage{tikzscale}
\usepackage{babel}
\usepackage{standalone}
\usepackage{xcolor}
\usepackage{dcolumn}   
\usepackage{amssymb}   
\usepackage{mathtools,empheq}
\usepackage{hyperref}
\usepackage{booktabs}
\usepackage{float}
\usepackage{epstopdf}
\usepackage{siunitx} 
\usepackage{gensymb} 
\usepackage{braket}
\usepackage{bm} 
\usepackage{array} 
\newcolumntype{C}[1]{>{\centering\arraybackslash}m{#1}}
\newcommand{\et}{{\em et al.~}}

\newcommand{\h}[1]{}

\newcommand{\qquestion}[1]{}
\usepackage[capitalise]{cleveref} 
\Crefname{figure}{Figure}{Figures}

\hypersetup{
	colorlinks=true,
	linkcolor=blue,
	filecolor=blue,
	urlcolor=blue,
	citecolor=blue,
}
\newcolumntype{C}[1]{>{\centering\arraybackslash}m{#1}}
\usepackage{tikz}
\usepackage{pgfplots}

\begin{document}

\title{
    Convergent close-coupling approach to ion collisions with multi-electron
    targets: Application to $\bar{p} + {\rm C}$ collisions
}

\author{N.~W.~Antonio\,\orcidlink{0000-0003-3900-6197}}
\email{nicholas.antonio@postgrad.curtin.edu.au}
\affiliation{Department of Physics and Astronomy, 
    Curtin University, 
    GPO Box U1987, 
    Perth, 
    WA 6845, 
    Australia
}

\author{A.~S.~Kadyrov\,\orcidlink{0000-0002-5804-8811}}
\affiliation{Department of Physics and Astronomy, 
    Curtin University, 
    GPO Box U1987, 
    Perth, 
    WA 6845, 
    Australia
}
\affiliation{Institute of Nuclear Physics, 
    Ulugbek, 
    100214 Tashkent, 
    Uzbekistan
}

\date{\today}

\begin{abstract}
    The single-centre convergent close-coupling approach to ion-atom collisions 
    has been extended to model collisions involving arbitrary multi-electron 
    atoms and partially stripped ions. This is accomplished by generating a set 
    of target pseudostates using the configuration interaction method. The 
    resulting pseudostates are expanded in terms of configuration state 
    functions, constructed using a hybrid of Hartree-Fock and Coulomb-Sturmian 
    spin-orbitals. This new approach is applied to study antiproton collisions 
    with atomic carbon. We present excitation energies, oscillator strengths, 
    and the dipole polarisability obtained using the target structure model to 
    validate its accuracy. Furthermore, we present results for 
    elastic-scattering, total excitation, and ionisation cross sections in the 
    incident energy range between 10 to 1000 keV. State-resolved excitation 
    cross sections for the first few dominant transitions are also presented. 
    Throughout the manuscript, we compare results obtained using the multi-core 
    target structure model with those from a frozen-core one. In all cases, we 
    find that a multi-core description of the carbon atom target is essential 
    for accurately modelling these collisions.
\end{abstract}
	
\maketitle

\section{introduction}

Collisions between heavy ions and multi-electron atoms or ions occur in many 
practical settings. They take place within fusion reactors during their 
operation \cite{Hill_2023_63}, in astrophysical plasmas \cite{Gu_2023_00} and 
during hadron therapy for cancer treatment \cite{Belkic_2010_47}. An accurate 
and comprehensive set of cross section data for these fundamental processes is 
essential for the continued development and optimization of such technologies. 
Hadron therapy uses high-energy heavy ions to deliver a concentrated dose of 
radiation to the tumour while minimizing damage to the surrounding healthy 
tissue \cite{Kraan_2015_05}. This contrasts with traditional X-ray therapy, 
which irradiates the tumour but deposits most of the dose into healthy tissue 
upon entry into the body. The key advantage of hadron therapy lies in the fact 
that heavy ions deposit most of their energy in the region of the Bragg peak 
towards the end of their range. The depth at which the Bragg peak occurs can be 
tuned, allowing clinicians to target tumours located at different depths within 
the body. To accurately configure where the Bragg peak occurs, and therefore 
the region where the maximum radiation dose is deposited, extensive treatment 
planning including depth-dose simulations is required \cite{Kraan_2015_05}. 
The accuracy of these simulations is paramount in ensuring the effectiveness 
of the treatment. However, the primary limiting factor in their accuracy is the 
lack of accurate cross section data for the underlying ion collisions. 
Therefore, both calculating and measuring these cross sections is of utmost 
importance to the continued advancement of hadron therapy.

There are many different types of ions that can be used as beams for
hadron therapy, and there is active research into various candidate ions for
future applications \cite{Rossi_2022_04}. In fact, proton and carbon ion 
therapy are currently in use for treatment today. One particularly interesting 
candidate ion that has been proposed for use in hadron therapy is the 
antiproton \cite{Knudsen_2008_266}. Antiprotons exhibit an additional effect, 
compared to positive ions, due to their annihilation at the end of their range. 
This annihilation process results in approximately 2.75 times more energy being 
deposited at the tumour site compared to proton and carbon ions 
\cite{Knudsen_2008_266}. Despite these promising benefits of antiproton 
therapy, it remains in the experimental phase. However, research into the use 
of antiprotons in hadron therapy is ongoing 
\cite{Dimcovski_2025_00,Chattaraj_2022_67}. 

In reality, ions used in hadron therapy collide with complex molecules within 
the human body, such as DNA and proteins, which are composed of many atoms. 
Modelling collisions with such targets in a completely ab initio manner is 
currently infeasible due to the complexity of the problem. However, current 
approaches to effectively model such collisions involve using the so-called 
independent atom model (IAM) \cite{Ludde_2022_106}. Applying the IAM to compute 
cross sections relies on accurate calculations of ion collisions with the 
corresponding atoms that constitute the molecular targets. One of the most 
important biologically relevant atomic targets is carbon, as it is a primary
constituent of organic molecules, and is therefore present in all biological 
systems.

Significant progress has been made in the past few decades on modelling ion 
collisions with one- and two-electron targets, in particular. A number of 
different theoretical approaches currently exist to model such collisions. 
These include the classical trajectory Monte Carlo (CTMC) \cite{Olson_1977_16}, 
continuum distorted wave (CDW) \cite{Cheshire_1964_84}, molecular orbital 
close-coupling (MOCC) \cite{Kimura_1989_00}, atomic orbital close-coupling 
(AOCC) \cite{Winter_1986_33}, and lattice numerical solution of the 
time-dependent Schr\"odinger equation (LTDSE) \cite{Schultz_1999_82} methods. 
For recent reviews on the subject of ion-atom collisions, see Refs. 
\cite{Belkic_2019_00,Schultz_2019,Tribedi_2024_00}. The combination of these 
methods along with experimental measurements has provided a wealth of data for 
ion collisions with one- and two-electron targets. However, the situation is 
quite different for multi-electron targets, where accurate data is still quite 
limited. The primary challenge lies in accounting for the electron-electron 
correlation effects, and the growing number of interactions that need to be 
accounted for as the number of electrons in the target increases. 

With recent advancements in computing resources \cite{Abdurakhmanov_2025_00}, 
it is now becoming feasible to model ion collisions with multi-electron targets 
with sufficient accuracy. A notable method developed to study heavy ion 
collisions with multi-electron targets is the basis generator method (BGM) 
\cite{Ludde_2021_104}. This approach was applied to study the energy loss in 
antiproton collisions with helium, carbon, nitrogen, oxygen and neon atoms in 
Ref. \cite{Ludde_2021_104}. They performed two sets of calculations, one where 
the model potential which accounts for the electron-electron interactions is 
independent of the position of the projectile ion (referred to as the 
no-response model), and another where this potential is dependent on the 
position of the projectile ion (referred to as the response model) 
\cite{Kirchner_2000_62}. The results displayed good agreement with experimental 
data for the He and Ne atoms. However at the same time, no experimental or 
theoretical data existed for the C, N, and O atoms to compare with. Another 
recently developed method to study antiproton collisions with Ne and Ar atoms 
is a time-dependent close-coupling approach by \citet{Jia_2024_110}. This 
approach used a pseudostate basis of configuration-interaction (CI) 
wavefunctions built from gaussian-type orbitals. 

Various quantum-mechanical (QM) and semi-classical (SC) implementations of the 
convergent close-coupling (CCC) approach to ion-atom collisions, using an 
orthogonal Laguerre, wave-packet (WP), and Coulomb-Sturmian (CS) bases have been 
applied to a wide range of collisions involving one- and multi-electron targets 
including H 
\cite{Abdurakhmanov_2011_44,Avazbaev_2016_93,Abdurakhmanov_2016_49,Abdurakhmanov_2018_97,Antonio_2024_110}, 
\ce{He} \cite{Alladustov_2019_99,Spicer_2024_109}, the inert gases 
\cite{Abdurakhmanov_2015_91}, the alkalis 
\cite{Abdurakhmanov_2020_53,Abdurakhmanov_2021_104}, and \ce{H_2} 
\cite{Abdurakhmanov_2013_111,Abdurakhmanov_2014_89,Plowman_2022_76}.
In general, the CCC framework, both in QM and SC formulations, has been able to 
reproduce experimental measurements with high accuracy. In this work we extend 
the single-centre Coulomb-Sturmian convergent close-coupling (CS-CCC) approach 
to ion-atom collisions to model collisions with arbitrary multi-electron 
targets without the need to use an effective one-electron approach. To this end 
we develop a new general atomic structure code based on the CI method that is 
capable of representing the entire spectra of any atom using the $LS$ coupling 
scheme. Having this code developed in-house allows us to seamlessly integrate 
it with our existing CS-CCC scattering code to maximise the computational 
performance of our calculations. We use a hybrid set of spin-orbitals 
consisting of Hartree-Fock (HF) \cite{Fischer_2019_00} and Coulomb-Sturmian 
\cite{Rotenberg_1970_00} ones. These multi-electron CI states are then 
used to expand the total scattering wavefunction, which upon substituting into 
the full Schr\"odinger equation, gives us a set of coupled-channel equations 
for the transition probability amplitudes. To demonstrate the capabilities of 
this new approach, we apply it to study $\bar{p} + {\rm C}$ collisions. This is 
motivated not only for its relevance to hadron therapy, but also because it is 
one of the first atoms in the periodic table that requires core-excited 
electron configurations to build an accurate target structure model. This 
makes it a perfect test case for the new approach and corresponding code. We 
will present calculations of the total elastic-scattering, excitation, and 
ionisation cross sections for these collisions in the incident energy range 
from 10 to 1000 keV. Furthermore, we will present results for some of the 
dominant state-resolved excitation cross sections for transitions from the 
$2s$ and $2p$ shells.

The remainder of this manuscript is structured as follows. In Sec. 
\ref{sec:theory} we present our theoretical approach to model 
$\bar{p} + {\rm C}$ collisions. In Sec. \ref{sec:details} we present
details of our calculations, including details of the target structure
model and numerical parameters required to obtain converged results. In Sec. 
\ref{sec:results} we present the results of our calculations, and 
finally in Sec. \ref{sec:conclusions} we present our conclusions and 
outlook for future work.

Atomic units are used throughout this work unless otherwise stated.

\section{Theory}\label{sec:theory}
Details of the single-centre semi-classical CCC approach to ion-atom collisions 
have been discussed in previous works, see e.g. \cite{Abdurakhmanov_2016_94,
Antonio_2025_111}. Therefore, a brief overview of the scattering part of the 
theory will be presented here. The focus of this section will be on 
incorporation of the multi-electron atomic structure model into the CS-CCC 
approach. Furthermore, as we are interested in collisions in the incident 
energy range between 10 to 1000 keV, we describe $\bar{p} + {\rm C}$ collisions 
using the single-centre approach.

\subsection{Target structure}

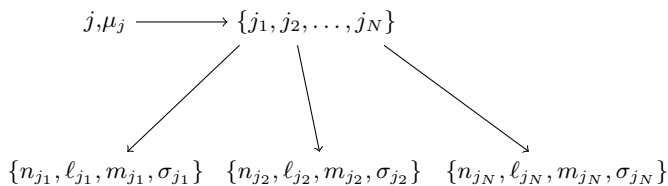
\begin{figure} [H]
    \begin{tikzpicture}
        \node (j) at (6, 2) {$j$,$\mu_{j}$};
        \node (j1j2) at (8.8, 2) {$\{j_1, j_2, \ldots, j_N\}$};
        \node (kj1) at (6.0, 0) {$\{n_{j_1}, \ell_{j_1}, m_{j_1}, \sigma_{j_1}\}$};
        \node (kj2) at (8.9, 0) {$\{n_{j_2}, \ell_{j_2}, m_{j_2}, \sigma_{j_2}\}$};
        \node (kjN) at (12, 0) {$\{n_{j_N}, \ell_{j_N}, m_{j_N}, \sigma_{j_N}\}$};

        \draw[->] (j) -- (j1j2);
        \draw[->] (7.78,1.69) -- (kj1);
        \draw[->] (8.55,1.69) -- (kj2);
        \draw[->] (9.7,1.69) -- (kjN);
    \end{tikzpicture}
    \caption{Schematic diagram of the indexing used to label the $N$-electron
    configurations.}
    \label{indexes}
\end{figure}

For generality, let us consider some arbitrary $N$-electron target atom or ion 
with a corresponding Hamiltonian $H_{\rm T}$, given by
\begin{align}
    H_{\rm T} = \sum_{i=1}^{N} \left(-\frac{1}{2}\nabla_i^2 - \frac{Z_{\rm 
    T}}{r_i}\right)
    + \sum_{i<j}^{N} \frac{1}{|\bm{r}_i - \bm{r}_{j}|},
\end{align}
where $Z_{\rm T}$ is the charge of the target nucleus, and $\bm{r}_i$ 
($\bm{r}_{j}$) is the position of the $i$-th ($j$-th) electron relative to
the nucleus. We construct each $N$-electron pseudostate $\ket{\psi_{\alpha}}$ 
of the target using the CI method. Furthermore, we use the $LS$-coupling scheme 
to label the atomic states. This way, the good quantum numbers needed to 
describe the states are the total orbital angular momentum $L$, its projection 
$M$, the total spin $S$, its projection $\Sigma$ and the parity $\Pi$. 
Including an index $n$ to distinguish between states with the same set of 
quantum numbers, we use the label $\alpha$ to represent the following set 
$\{n,L,M,S,\Sigma,\Pi\}$ for the pseudostate $\ket{\psi_{\alpha}}$. Each 
pseudostate is expanded in terms of a set of configuration state functions 
(CSFs) represented by $\ket{j:L,M,S,\Sigma,\Pi}$ as follows
\begin{align}
    \ket{\psi_{\alpha}} = \sum_{j=1}^{M_{LS\Pi}}
    c^{(\alpha)}_{j} 
    \ket{j:L,M,S,\Sigma,\Pi},
\end{align}
where $M_{LS\Pi}$ is the number of CSFs used to expand the pseudostate
$\ket{\psi_{\alpha}}$, which differs depending on the quantum numbers
$L$, $S$ and $\Pi$. For the remainder of this work, for brevity, we may often 
omit the magnetic quantum numbers $M$ and $\Sigma$ from the notation of the 
CSFs, and assume that they are implicitly present. The CI coefficients, denoted 
as $c^{(\alpha)}_{j}$, are determined by diagonalising the target Hamiltonian 
$H_{\rm T}$ in the space spanned by the chosen set of CSFs. Index $j$ 
represents a single electron configuration 
$\{(n_{1},l_{1}),(n_{2},l_{2}),\ldots,(n_{N},l_{N})\}$, 
as well as a unique set 
of intermediate orbital and spin angular momentum couplings.
The CSFs are constructed in such a way that they are eigenstates of $\hat{L}^2$ 
and $\hat{S}^2$, where $\hat L$ and $\hat S$ are the operators of the total 
orbital angular momentum and total spin, respectively, along with the 
projection operators $\hat{L}_z$ and $\hat{S}_z$. This is achieved by writing 
each CSF as a linear combination of Slater determinants 
$\ket{\chi_{j_1},\ldots,\chi_{j_N}}$ with appropriate coefficients as follows
\begin{align}
    \ket{j:L,S,\Pi} = \sum_{\mu_{j}\in\Gamma_{j}}
    A^{(L,S,\Pi)}_{\mu_{j}}
    \ket{\chi_{j_1},\ldots,\chi_{j_N}},
    \label{csf} 
\end{align}
where $\mu_{j}$ is a set of one-electron projection quantum numbers
$\{(m_{1},\sigma_{1}),(m_{2},\sigma_{2}),\ldots,(m_{N},\sigma_{N})\}$
and $\Gamma_{j}$ is the collection of all possible sets of one-electron
projection quantum numbers for the $j$-th electron configuration. For clarity,
we collect all four one-electron quantum numbers for the $i$-th electron into a 
single index $j_i$. To help articulate the index notation we use, refer to a 
schematic diagram in Fig. \ref{indexes} which illustrates the mappings between 
the various indices used to define electron configurations. The coefficients 
$A^{(L,S,\Pi)}_{\mu_{j}}$ are the angular momentum coupling ones which 
are determined by coupling each of the $N$-electron orbital angular momenta 
$\ell_{j_i}$ and spins $\sigma_{j_i}$, along with their associated projection
quantum numbers.

The Slater determinants can be written as antisymmetrised products of 
one-electron spin-orbitals $\ket{\chi_{j_i}}$ as follows
\begin{align}
    \ket{\chi_{j_1},\chi_{j_2},\ldots,\chi_{j_N}}
    = \frac{1}{\sqrt{N!}}\mathcal{A}
    \ket{\chi_{j_1}}\ket{\chi_{j_2}}\ldots\ket{\chi_{j_N}},
\end{align}
where $\mathcal{A}$ is the antisymmetrisation operator. 

Defining $\bm{x}$ as the position and spin of the $i$-th electron, we can write 
the spin-orbital in the position representation as follows
\begin{align}
    \chi_{j_i}(\bm{x})
    &=\frac{1}{r}
    \varphi_{n_{i}\ell_{i}}(r)Y_{\ell_{i}m_{i}}(\hat{\bm{r}})
    X_{\frac{1}{2}}(\sigma_{i}),
\end{align}
where in general $\varphi_{n_{i}\ell_{i}}(r)$ is some square-integrable 
radial function, $Y_{\ell_{i}m_{i}}(\hat{\bm{r}})$ is spherical harmonics,
and $X_{\frac{1}{2}}(\sigma_{i})$ is a spin function. It is worth mentioning 
that the choice of the radial functions $\varphi_{n_{j_i}\ell_{j_i}}(r)$ is 
very important as it strongly influences the convergence rate of the CI 
calculations with increasing the size of the expansion ($M_{LS\Pi}$). 
To best describe the low-lying states of the target atom, while also 
maintaining adequate representation of the continuum, we use a hybrid set of
radial functions consisting of HF and CS ones. The HF radial functions are
generated using the program by \citet{Froese_Fischer_1987_43}. The present
structure code is written in such a way that the spin-orbitals do not need to 
be orthogonal to each other, allowing for further flexibility in the choice of
the radial functions.

The CI coefficients $c_{j}^{(\alpha)}$ and associated energy eigenvalues
$\varepsilon_{\alpha}$ can be obtained by solving the following generalized
eigenvalue problem for each $L$, $S$ and $\Pi$ needed for convergence of 
the subsequent scattering calculations:
\begin{align}
    \sum_{j=1}^{M_{LS\Pi}}
    \Big( 
        H^{(L,S,\Pi)}_{ij} -
        \varepsilon_{\alpha}
        S^{(L,S,\Pi)}_{ij}
    \Big)
    c_{i}^{(\alpha)} = 0,
\end{align}
where $H^{(L,S,\Pi)}_{ij}$ and $S^{(L,S,\Pi)}_{ij}$ are the Hamiltonian and
overlap matrix elements between the CSFs which can be written in terms of 
matrix elements between the Slater determinants as follows
\begin{align}
    S^{(L,S,\Pi)}_{ij} &= 
    \sum_{\mu_{i}\mu_{j}}
    A^{(L,S,\Pi)}_{\mu_{i}}A^{(L,S,\Pi)}_{\mu_{j}}
    \braket{\chi_{i_1},\ldots,\chi_{i_N}|\chi_{j_1},\ldots,\chi_{j_N}},
    \nonumber \\
    H^{(L,S,\Pi)}_{ij} &=
    \sum_{\mu_{i}\mu_{j}}
    A^{(L,S,\Pi)}_{\mu_{i}}A^{(L,S,\Pi)}_{\mu_{j}}
    \braket{\chi_{i_1},\ldots,\chi_{i_N}|H_{\rm T}|\chi_{j_1},\ldots,\chi_{j_N}}.
    \label{HSmatrix}
\end{align}
We evaluate the matrix elements between various Slater determinants using the
generalized Slater-Condon rules \cite{Avery_2006_00}.

\subsection{Semi-classical convergent close-coupling approach}
Let us consider a collision between an ion with charge $Z_{\rm P}$, and an 
$N$-electron target atom. The Hamiltonian for the collision system can be 
written as
\begin{align}
    H = K_{\bm{\sigma}} + H_{\rm T} + \overline{V},
\end{align}
where $K_{\bm{\sigma}}$ is the kinetic energy operator for the relative motion
of the incoming antiproton projectile and the target atom, making $\bm{\sigma}$
the position of the antiproton relative to the target atom. The operator 
$\overline{V}$ represents the interaction potential between the antiproton and 
the target atom which can be written as
\begin{align}
    \overline{V} = 
    \frac{Z_{\rm P}Z_{\rm T}}{R} 
    - Z_{\rm P}
    \sum_{i=1}^{N} \frac{1}{|\bm{r}_{i} - \bm{R}|},
\end{align}
where $\bm{R}$ is the position of the antiproton relative to the target nucleus.
As the Hamiltonian of the collision system is independent of the electron 
spin, it is trivial to show that the total spin of the system is conserved.
Therefore, for the remainder of this section for brevity, we drop the initial
spin of the target $S$ and its projection being $\Sigma$
from the notation. We wish to find 
solutions to the full Schr\"odinger equation of the collision system
\begin{align}
    (H - E) \Psi^{+}_{i}(\bm{\sigma},\bm{x}_{1},\ldots,\bm{x}_{N})=0,
    \label{SE}
\end{align}
where $E$ is the total energy of the collision system, superscript $+$ indicates
the scattering wavefunction has outgoing-wave boundary conditions, and $i$
specifies the initial state of the target atom. In order to solve Eq. (\ref{SE}),
we expand the total scattering wavefunction $\Psi^{+}_{i}$ in terms of $N_{\rm 
T}$ target pseudostates $\psi_{\alpha}$ as follows
\begin{align}
    \Psi^{+}_{i}(\bm{\sigma},\bm{x}_{1},\ldots,\bm{x}_{N}) \approx
    \sum_{\alpha=1}^{N_{\rm T}}
    \mathcal{F}_{\alpha}(\bm{\sigma}) {\rm e}^{\bm{k}_{\alpha}\cdot \bm{\sigma}}
    \psi_{\alpha}(\bm{x}_{1},\ldots,\bm{x}_{N}),
    \label{scat_wf}
\end{align}
where $\bm{k}_{\alpha}$ is the relative momentum between the antiproton and the
target atom, and $\mathcal{F}_{\alpha}(\bm{\sigma})$ are initially unknown
expansion coefficients.

Inserting the expansion for the $\Psi^{+}_{i}$ into Eq. (\ref{SE}) and 
projecting it by $N_{\rm T}$ asymptotic states,
${\rm e}^{i\bm{k}_{\alpha'}\cdot \bm{\sigma}}\psi_{\alpha'}$, we are able to
obtain a set of coupled-channel equations for the unknown coefficients. 
To make the set numerically tractable, we employ the semi-classical 
approximation. This approximation, models the motion of the incoming antiproton 
using a classical trajectory, while leaving the electron dynamics described 
fully quantum-mechanically. This allows us to write 
$\bm{R} \equiv \bm{R}(t,\bm{b}) = \bm{b} + \bm{v}t$, where $\bm{b}$ is the 
impact parameter, $\bm{v}$ is the initial velocity of the antiproton, and $t$ 
is the time. The time parameterisation is defined such that $t= - \infty$ 
corresponds to the moment prior to the collision, $t=0$ corresponds to the 
moment of the closest approach, and $t=+\infty$ corresponds to time after the 
collision. Next we Taylor expand the coefficients $\mathcal{F}_{\alpha}$ about 
$\bm{R}$. Safely assuming that second and higher order terms in the expansion 
are negligible, we can write the coefficients as functions of $\bm{R}$, like 
$\mathcal{F}_{\alpha}(\bm{\sigma}) \approx F_{\alpha}(t,\bm{b})$. Next,
the semi-classical approximation is applied, reducing the coupled-channel 
equations to
\begin{align}
    i \frac{{\rm d}F_{\alpha'}(t,\bm{b})}{{\rm d}t} 
    &=
    \sum_{\alpha=1}^{N_{\rm T}}
    F_{\alpha}(t,\bm{b})\braket{\psi_{\alpha'}|\overline{V}|\psi_{\alpha}}
    {\rm e}^{iv(k_{\alpha} - k_{\alpha'})t},
    \label{cceqs}
\end{align}
where the index $\alpha'$ runs from $1$ to $N_{\rm T}$. These close-coupling 
(CC) equations are solved with the initial condition
\begin{align}
    F_{\alpha}(t=-\infty,\bm{b}) = \delta_{\alpha i},
\end{align}
for all $\alpha$ from $1$ to $N_{\rm T}$. In this work, $i$ is set to 
correspond to the ground \ce{$1s^22s^22p^2$ ^3P^{e}} state of the
carbon atom with different initial total magnetic quantum numbers $M=0, \pm 1$.

In order to solve the set of Eqs. (\ref{cceqs}), we need to first 
evaluate the scattering matrix elements 
$\braket{\psi_{\alpha'}|\overline{V}|\psi_{\alpha}}$. To do 
this, we first expand them in terms of matrix elements between
the CSFs $\braket{i: L', M', \Pi' || \overline{V} || j: L, M, \Pi}$
\footnote{
    Here, we explicitly show the magnetic quantum numbers $M'$ and $M$ of the
    CSFs to emphasise that these matrix elements are dependent on these
    numbers. 
}
, as follows
\begin{align}
    \braket{\psi_{\alpha'}|\overline{V}|\psi_{\alpha}} 
    &=
    \sum_{ij}
    c^{(\alpha')}_{i}c^{(\alpha)}_{j}
    \braket{i: L', M', \Pi' | \overline{V} | j: L, M, \Pi}.
    \label{cceqs2}
\end{align}
Next, we partial-wave expand the interaction potential as follows
\begin{align}
    \overline{V}(\bm{R},\bm{r}_{1},\ldots,\bm{r}_{N}) =
    \sum_{\lambda \mu}
    \overline{V}^{(\lambda)}_{\mu}(R,\bm{r}_{1},\ldots,\bm{r}_{N})
    C^{(\lambda)*}_{\mu}(\hat{\bm{R}}),
    \label{Vexpansion}
\end{align}
where $C^{(\lambda)}_{\mu}(\hat{\bm{R}})$ are the renormalised spherical 
harmonics as defined in Ref. \cite{Fischer_2019_00} and
$\overline{V}^{(\lambda)}_{\mu}(R,\bm{r}_{1},\ldots,\bm{r}_{N})$ is 
given by
\begin{align}
    \overline{V}^{(\lambda)}_{\mu}(R,\bm{r}_{1},\ldots,\bm{r}_{N}) =
    Z_{\rm P}
\sum_{k=1}^{N}\mathcal{U}_{\lambda}(r_{k},R)Y_{\lambda \mu}(\hat{\bm{r}}_{k}).
\label{Vlambda}
\end{align}
The functions $\mathcal{U}_{\lambda}(r_{k},R)$ are written as
\begin{align}
    \mathcal{U}_{\lambda}(r_{k},R) = 
    \begin{cases}
        \frac{Z_{\rm T}}{R}\delta_{\lambda 0}-
        \frac{r_{k}^{\lambda}}{R^{\lambda+1}}, & \text{if } r_{k} \leq R, \\
        \frac{Z_{\rm T}}{R}\delta_{\lambda 0}-
        \frac{R^{\lambda}}{r_{k}^{\lambda+1}}, & \text{if } r_{k} > R.
    \end{cases}
\end{align}
Substituting the expansion given in Eq. (\ref{Vexpansion}) and applying the 
Wigner-Eckart theorem,  the matrix elements 
in Eq. (\ref{cceqs}) reduce to
\begin{widetext}
\begin{align}
    \braket{\psi_{\alpha'}|\overline{V}|\psi_{\alpha}} 
    &=
    \sum_{ij\lambda}
    c^{(\alpha')}_{i}c^{(\alpha)}_{j}
    C^{L',M'}_{L,M,\lambda,\mu}
    \braket{i: L', \Pi' || \overline{V}^{(\lambda)}_{\mu} || j: L, \Pi}
    C^{(\lambda)*}_{\mu}(\hat{\bm{R}}),
    \label{final_dmat}
\end{align}
\end{widetext}
where $C^{L',M'}_{L,M,\lambda,\mu}$ is a Clebsch-Gordan coefficient, and
$\braket{i: L', \Pi' || \overline{V}^{(\lambda)}_{\mu} || j: L, \Pi}$ is a
reduced matrix element, independent of the magnetic quantum numbers, 
for the transition between the two CSFs.

To evaluate the reduced matrix elements in Eq. (\ref{final_dmat}), we first 
evaluate the corresponding matrix elements that depend on the magnetic quantum 
numbers, 
$\braket{i: L', M', \Pi' | \overline{V}^{(\lambda)}_{\mu} | j: L, M, \Pi}$, for 
any given values of $M'$ and $M$. The best choice of the magnetic quantum numbers to 
use to evaluate these matrix elements is $M' = L'$ and $M = L$, as these are 
the ones which minimise the number of Slater determinants needed to construct 
each corresponding CSF. The evaluation of the full matrix elements is done 
similarly to what is done for the Hamiltonian and overlap matrix elements in 
Eq. (\ref{HSmatrix}). Specifically, we first expand the CSFs in terms of their 
respective Slater determinants to get
\begin{align}
    &\braket{i: L', M', \Pi' | \overline{V}^{(\lambda)}_{\mu} | j: L, M, \Pi}
    =
    \nonumber \\
    &\quad\sum_{\mu_{i}\mu_{j}}
    A^{(L',S,\Pi')}_{\mu_{i}}A^{(L,S,\Pi)}_{\mu_{j}}
    \braket{\chi_{j_1},\ldots,\chi_{j_N}|\overline{V}^{(\lambda)}_{\mu}|
    \chi_{j_1},\ldots,\chi_{j_N}}.
    \label{full_dmat}
\end{align}
After applying the generalised Slater-Condon rules to determine the matrix 
elements between the Slater determinants and using Eq. (\ref{Vlambda}), we get 
the final expression for the full matrix elements in terms of the one-electron 
matrix elements as follows
\begin{widetext}
\begin{align}
    \braket{i: L', M', \Pi' | \overline{V}^{(\lambda)}_{\mu} | j: L, M, \Pi}
    &= 
    Z_{\rm P}
    \sum_{\mu_{i}\mu_{j}ab}
    (-1)^{(a+b)}
    A^{(L',S,\Pi')}_{\mu_{i}}A^{(L,S,\Pi)}_{\mu_{j}}
    \braket{\chi_{i_a}|\mathcal{U}_{\lambda}(r,R)|\chi_{j_b}}
    |S(i,j|a,b)|, 
\label{full_dmat2}
\end{align}
\end{widetext}
where $\braket{\chi_{i_a}|\mathcal{U}_{\lambda}(r,R)|\chi_{j_b}}$ is the 
one-electron matrix element between the spin-orbitals $\chi_{i_a}$ and 
$\chi_{j_b}$ which can be evaluated as shown in previous works such as in Ref. 
\cite{Abdurakhmanov_2011_44}. The term $S(i,j|a,b)$ is the matrix of 
one-electron overlaps between all the spin orbitals from electron 
configuration $i$ and $j$ with row $a$ and column $b$ removed, making 
$|S(i,j|a,b)|$ its determinant. With these matrix elements known, we can then
apply an inverse application of the Wigner-Eckart theorem to obtain the
reduced matrix elements 
$\braket{i: L', \Pi' || \overline{V}^{(\lambda)}_{\mu} || j: L, \Pi}$. As 
evaluating Eq. (\ref{full_dmat2}) is computationally expensive, following the 
procedure outlined above we are able to drastically speed up the calculations 
by only needing to calculate magnetic quantum number-dependent matrix
elements for only one pair of $M'$ and $M$ for each CSF pair.

\subsection{Cross sections}
It has been shown in Ref. \cite{Abdurakhmanov_2020_53}, that in the limit of
$t\to +\infty$, the expansion coefficients $F_{\alpha}(+\infty,\bm{b})$
represent the transition probability amplitudes in the impact parameter
representation. As a result, the probability of transition from some initial 
state $i$ to some final state $f$ can be calculated as follows
\begin{align}
    P^{(M_f,M_i)}_{fi}(b) 
    = |F_{f}(+\infty,\bm{b})-\delta_{f i}|^2,
\end{align}
where the superscript $(M_f,M_i)$ indicates that the transition
probabilities are fully state-resolved, i.e. they depend on the
magnetic quantum numbers $M_f$ and $M_i$ of the final and initial states,
respectively. The corresponding total cross section for the transition can be 
calculated by integrating the transition probabilities with respect to the
parameter $\bm{b}$ as follows
\begin{align}
    \sigma^{(M_f,M_i)}_{fi} = 2\pi
    \int_{0}^{+\infty} bP^{(M_f,M_i)}_{fi}(b) {\rm d}b.
    \label{cross_section}
\end{align}
If the initial state of the target atom has $L_{i} > 0$, then there are
$2L_{i}+1$ degenerate states with the same energy. As a result, we can
define excitation cross sections independent of the magnetic quantum
numbers by averaging over all the initial-state magnetic quantum numbers
and summing over the final ones as follows
\begin{align}
    \sigma_{fi} = \frac{1}{2L_{i}+1}
    \sum_{M_{i}M_{f}} \sigma^{(M_f,M_i)}_{fi}.
    \label{cross_section_avg}
\end{align}
All results presented in this work are averaged over the initial state magnetic
quantum numbers according to Eq. (\ref{cross_section_avg}).

\section{Details of the calculations}\label{sec:details}

\subsection{Target structure model}
Before scattering calculations can be performed, we need to first construct a 
atomic structure model which can adequately represent the spectrum of the carbon 
atom target. To begin with, we performed HF calculations for the ground state 
of the carbon atom. This produced a set of radial orbitals corresponding to the 
$1s$, $2s$ and $2p$ subshells. The remaining spin orbitals were constructed 
using the CS radial functions, with falloff parameter set to 1.0. To account 
for the correlations between the valence electrons, we included configurations 
of the form $1s^2 2s^2 n\ell n'\ell'$. Depending on the term symbol 
$^{2S+1}L^{\Pi}$ the range of $n$, $\ell$, and $\ell'$ were different. In 
summary, across all symmetries considered, $n$ went up to at most 9 and $\ell$ 
went up to at most $3$ depending on the term symbol of the configuration, 
while $n'$ went up to 18 in all cases and $\ell'$ went up to at most $4$. 
From here on, we define the maximum value of $n'$ used in the model as 
$n_{\rm max}$. In addition, to account for the correlations between the valence 
and core electrons, we included the following set of electron configurations: 
$1s^2 2p^2 n\ell n'\ell'$, $1s^2 2s 2p n\ell n'\ell'$ and $1s^2 2s (n\ell)^3$. 
Just as for the valence-valence electron configurations, in all cases $n'$ went 
up to $n_{\rm max}$ and the range of $n$, $\ell$ and $\ell'$ depended on the 
term symbol and specific type of electron configuration with the same range as 
specified above.

Throughout the manuscript we compare results obtained using a frozen-core (FC)
and multi-core (MC) description of the carbon atom. This is to emphasise the 
necessity to describe the C structure using the MC description, as well as the 
importance of the core-valence electron correlations. The FC approach only 
includes electron configurations of the following form: $1s^2 2s^2 2p n\ell$, 
where the maximum value of $n$ was $n_{\rm max}$ and $\ell$ went up to at most 
$4$. The excitation energies of the first few states of C obtained using both 
the FC and MC approaches is presented in Table \ref{tab:excitation_energies}. 
Alongside these energies we also show the ones given in the NIST database 
\cite{NIST_ASD}. The last column shows the percentage difference between the 
present MC results and the NIST data. Firstly, we note that the largest 
difference between the present FC and MC calculations, which is 8\%, appears to 
be for the $2s^2 2p3p$ \ce{^3P^{e}} state. Another interesting observation is 
that while the energies of the $2s^2 2p 3p$ \ce{^3P^{e}}, 
$2s^2 2p 3p$ \ce{^3S^{e}} and $2s^2 2p 3p$ \ce{^3D^{e}} states in the MC model 
are almost degenerate, they still appear to require the multi-core CSFs to be 
accurately described. This is seen by comparing their respective energies in 
the FC and MC calculations. The largest difference between the NIST data and 
these first several states of the atom obtained in the MC model is 2.53\% and 
that is for the $2s^2 2p 4s$ \ce{^3P^{o}} state. However, for modelling 
$\bar{p} + \ce{C}$ collisions, this level of accuracy is acceptable.

When working on atomic collision problems with methods such as coupled-channel 
ones, it is not feasible to reach the level of precision in the structure 
models similar to that required in quantum chemistry studies 
\cite{Nasiri_2024_122}. This is because we need to balance the accuracy of the 
first few bound states while still being able to adequately represent the 
continuum using a reasonable number of pseudostates. To this effect, in Table 
\ref{tab:excitation_energies_comparison} we present a comparison between the 
present CI structure model excitation energies obtained using the MC approach 
and other CI calculations used in other atomic collision theory works. It is 
generally considered that the model of C by \citet{Wang_2013_87}, used to study
$e^{-} + {\rm C}$ collisions, is the most accurate one to compare to in the 
context of modelling collisions involving the carbon atom 
\cite{Stancalie_2015_576}. It was generated using the B-spline 
R-matrix codes by \citet{Zatsarinny_2006_174}. One of the main strengths of the 
Zatsarinny code is that it is capable of using term-dependent sets of spin 
orbitals. This means that in their approach, the radial functions can be 
optimised for each term symbol. The functions used in Ref. \cite{Wang_2013_87} 
was comprised of a set of multi-configurational Hartree-Fock (MCHF) radial ones 
for the C atom and B-spline functions. It is not explicitly stated how the set 
of radial functions used by \citet{Stancalie_2015_576} was generated. 
Generally, we find our excitation energies align closer to the benchmark one 
by \citet{Wang_2013_87} than the \citet{Stancalie_2015_576}.

\begin{table*}[t]
    \centering
    \renewcommand{\arraystretch}{1.5}
    \setlength{\tabcolsep}{12pt}
    \begin{tabular}{|c|c|c|c|c|c|}
        \hline
        \hline
        State & Term & Present (FC) & Present (MC) & NIST \cite{NIST_ASD} & \% diff \\
        \hline
        \ce{$2s^2 2p^2$}  & \ce{^3 P^{e}} & 0.0000 & 0.0000 & 0.0000 & 0.000 \\
        \ce{$2s^2 2p 3s$} & \ce{^3 P^{o}} & 0.2951 & 0.2758 & 0.2749 & 0.33 \\
        \ce{$2s 2p^3$}    & \ce{^3 D^{o}} & -      & 0.2937 & 0.2920 & 0.58 \\
        \ce{$2s^2 2p 3p$} & \ce{^3 D^{e}} & 0.3342 & 0.3235 & 0.3175 & 1.89 \\
        \ce{$2s^2 2p 3p$} & \ce{^3 S^{e}} & 0.3406 & 0.3236 & 0.3223 & 0.43 \\
        \ce{$2s^2 2p 3p$} & \ce{^3 P^{e}} & 0.3477 & 0.3235 & 0.3251 & 0.49 \\
        \ce{$2s 2p^3$}    & \ce{^3 P^{o}} & -      & 0.3429 & 0.3429 & 0.00 \\
        \ce{$2s^2 2p 4s$} & \ce{^3 P^{o}} & 0.3750 & 0.3649 & 0.3559 & 2.53 \\
        \hline
        Ionis. Limit & & 0.4118 & 0.4172 & 0.4138 & 0.82 \\
        \hline
        \ce{$2s 2p^3$}    & \ce{^3 S^{o}} & -      & 0.4765 & 0.4821 & 1.16 \\
        \hline
        \hline
    \end{tabular}
    \caption{Comparison of the present CI results for the excitation energies of
        the triplet states of the C atom with the NIST data. Both frozen-core
        (FC) and multi-configuration (MC) CI calculations are shown. The 
        last column shows the percentage difference between the present
        MC results and the NIST data. The ionisation limit is determined by 
        performing a CI calculation for the ground state of the \ce{C^{+}} ion
        and taking the difference between the ground state energy of the
    \ce{C^{+}} ion and the ground state energy of the \ce{C} atom.}
    \label{tab:excitation_energies}
\end{table*}

Aside from the set of excitation energies generated by the CI calculations, 
another crucial assessment of the quality of the structure model is the set of 
oscillator strengths it produces. In Table \ref{tab:oscillator_strengths}
we show the present oscillator strengths for transitions from the ground
state of the C atom to the first few excited states alongside other calculated 
values from the same works as in Table 
\ref{tab:excitation_energies_comparison}. Also shown are the corresponding NIST 
values \cite{NIST_ASD} for comparison. Generally, we find that our results 
agree well with the available NIST data, with the exceptions of the transitions 
to the \ce{$2s2p^24s$ ^3P^{o}} and \ce{$2s^22p3d$ ^3P^{o}} states. Excluding 
the previous two transitions, we also find our results agree well with the 
benchmark ones by \citet{Wang_2013_87}. However, the results by 
\citet{Stancalie_2015_576} are generally further away from the Wang \et results 
and our present calculations, as well as the NIST data. 

\begin{table*}[t]
    \centering
    \renewcommand{\arraystretch}{1.5}
    \setlength{\tabcolsep}{12pt}
    \begin{tabular}{|c|c|c|c|c|}
        \hline
        \hline
        State & Term & Present (MC) & \citet{Wang_2013_87} & 
        \citet{Stancalie_2015_576} \\
        \hline
        \ce{$2s^2 2p^2$}  & \ce{^3 P^{e}} & 0.0000 & 0.0000 & 0.0000  \\
        \ce{$2s^2 2p 3s$} & \ce{^3 P^{o}} & 0.2758 & 0.2766 & 0.2720  \\
        \ce{$2s 2p^3$}    & \ce{^3 D^{o}} & 0.2937 & 0.2941 & 0.3065  \\
        \ce{$2s^2 2p 3p$} & \ce{^3 D^{e}} & 0.3235 & 0.3178 & 0.3160  \\
        \ce{$2s^2 2p 3p$} & \ce{^3 S^{e}} & 0.3236 & 0.3225 & 0.3224  \\
        \ce{$2s^2 2p 3p$} & \ce{^3 P^{e}} & 0.3235 & 0.3255 & 0.3421  \\
        \ce{$2s 2p^3$}    & \ce{^3 P^{o}} & 0.3429 & 0.3446 & 0.3497  \\
        \ce{$2s^2 2p 4s$} & \ce{^3 P^{o}} & 0.3649 & 0.3554 & 0.3727  \\
        \hline
        Ionis. Limit & & 0.4172 & - &  \\
        \hline
        \ce{$2s 2p^3$}    & \ce{^3 S^{o}} & 0.4765 & 0.4804 & 0.4821 \\
        \hline
        \hline
    \end{tabular}
    \caption{Same as Table \ref{tab:excitation_energies} but the
        present MC CI results are compared with the structure calculations from 
        \citet{Wang_2013_87}, and \citet{Stancalie_2015_576}.}
    \label{tab:excitation_energies_comparison}
\end{table*}

The last quality check we perform on the present structure model is to compare 
the dipole polarisability of the ground state of the C atom with the
accepted value. Our model produces a value of 11.66 $a_0^{3}$, which is 
within 3\% of the accepted value of $11.3 \pm 0.2$ $a_0^{3}$ reported in Ref. 
\cite{Schwerdtfeger_2019_117}. Furthermore, this value is close to the one by 
\citet{Wang_2013_87} which is 11.31 $a_0^{3}$. It is important to emphasise 
that we are not only concerned with obtaining an accurate description of the 
bound state spectrum of the C atom, but also of the continuum. The obtained 
result for the dipole polarisability is an indication that our new structure 
model adequately represents the continuum in addition to the bound state 
spectrum.

With all of the above quality checks performed, we can be confident that the
present structure model accurately represents the spectrum of the C atom and 
can be used to model antiproton collisions with this atom.

\begin{table*}[t]
    \centering
    \renewcommand{\arraystretch}{1.5}
    \setlength{\tabcolsep}{12pt}
    \begin{tabular}{|c|c|c|c|c|}
        \hline
        \hline
        Transition & \citet{Wang_2013_87} & \citet{Stancalie_2015_576} & Present (MC) & NIST \cite{NIST_ASD} \\ 
        \hline
        \ce{$2s^2 2p^2$ ^3 P^{e}} $\to$ \ce{$2s^2 2p 3s$ ^3 P^{o}} & 0.143 & 0.124 & 0.130 & 0.140 \\ 
        \ce{$2s^2 2p^2$ ^3 P^{e}} $\to$ \ce{$2s 2p^3$ ^3 D^{o}}    & 0.073 & 0.098 & 0.071 & 0.072 \\ 
        \ce{$2s^2 2p^2$ ^3 P^{e}} $\to$ \ce{$2s 2p^3$ ^3 P^{o}}    & 0.056 & 0.028 & 0.069 & 0.063 \\ 
        \ce{$2s^2 2p^2$ ^3 P^{e}} $\to$ \ce{$2s 2p^24s$ ^3 P^{o}}  & 0.027 & 0.023 & 0.016 & 0.021 \\ 
        \ce{$2s^2 2p^2$ ^3 P^{e}} $\to$ \ce{$2s^2 2p 3d$ ^3 D^{o}} & 0.096 & 0.112 & 0.094 & 0.094 \\ 
        \ce{$2s^2 2p^2$ ^3 P^{e}} $\to$ \ce{$2s^2 2p 3d$ ^3 P^{o}} & 0.037 & 0.340 & 0.016 & 0.040 \\ 
        \ce{$2s^2 2p^2$ ^3 P^{e}} $\to$ \ce{$2s 2p^3$ ^3 S^{o}}    & 0.156 & 0.171 & 0.143 & 0.152 \\ 
        \hline
        \hline
    \end{tabular}
    \caption{The oscillator strengths for C. The present CI calculations are 
        shown alongside the results from \citet{Wang_2013_87} and
        \citet{Stancalie_2015_576}. The last
        column shows the data available from \cite{NIST_ASD}. 
    }
    \label{tab:oscillator_strengths}
\end{table*}

\subsection{Convergence tests and numerical parameters}
Aside from the accuracy of the atomic structure model, the accuracy of 
subsequent close-coupling calculations also depends on the level of 
convergence with respect to increasing the size of the expansion of the 
scattering wavefunction (\ref{scat_wf}).  In order to ensure that the
cross sections of interest converge to within a few percent, we perform 
the following series of convergence tests at three key incident energies:
10, 100, and 1000 keV. Firstly, we fix $n_{\rm max} = 18$ and 
establish convergence with respect to the number of symmetries (term symbols)
included in the coupled-channel calculations, i.e. used in the expansion of 
the scattering wavefunction (\ref{scat_wf}). To do this we begin with only 
including states with \ce{^3P^{e}} and \ce{^3P^{o}} symmetries. Then we 
continued to increase the total orbital angular momentum $L$, including both 
parities each time, until we reached $L=4$. Therefore, the final set of 
close-coupling calculations included states with the following term symbols:
\ce{^3P^{e}}, \ce{^3P^{o}}, \ce{^3D^{e}}, \ce{^3D^{o}}, \ce{^3S^{e}},
\ce{^3S^{o}}, \ce{^3F^{e}}, \ce{^3F^{o}}, \ce{^3G^{e}}, and \ce{^3G^{o}}. 
Having fixed the number of symmetries included in the calculations, we then 
established convergence with respect to increasing $n_{\rm max}$. Starting from 
$n_{\rm max} = 10$, we increased it by 2 until we reached $n_{\rm max} = 18$. 
The differences between the cross sections presented in this work using 
$n_{\rm max} = 18$ versus $n_{\rm max} = 16$ were less than 1\% in all cases. 

The CI calculations of the carbon atom produce a very dense set of pseudostates 
above the first ionisation threshold which represent its continuous spectrum. 
Even with today's computing resources, it is not feasible, nor necessary, to 
include all of them in subsequent close-coupling calculations. To reduce the 
number of pseudostates we do the following. We introduce a parameter denoted as
$\varepsilon_{\rm max}$ which specifies the maximum energy of the corresponding 
pseudostates to be included. We then increase this parameter until the
cross sections converge to the desired accuracy. We found that incrementing 
$\varepsilon_{\rm max}$ from 1.8 a.u. to 2.0 a.u. resulted in negligible
differences in the total and state-resolved excitation cross sections and a 
maximum of a 4\% difference in the ionisation cross section at 1000 keV.
This difference decreases to less than 1\% at 100 keV, and is negligible at
10 keV. In summary, our final CS-CCC calculations included 2590 pseudostates
with excitation energies up to 2.0 a.u. above the ground-state energy. This is 
approximately two times larger than the $\varepsilon_{\rm max}=1.03$ a.u. used 
by Wang \et in Ref. \cite{Wang_2013_87}. Note that the value of 
$\varepsilon_{\rm max}$ used by Wang \et, is below the double ionisation 
threshold. Therefore, their structure model only includes positive-energy 
states corresponding to single ionisation. Whereas the model used in the 
present work includes states corresponding to both single and double 
ionisation.

In order to keep the number of CSFs used to expand each of the pseudostates to 
a reasonable number, we also used a CI coefficient cutoff parameter. In all 
results presented in this work, we used a cutoff of 0.0005. This means that for 
each pseudostate $\psi_{\alpha}$, any CSF with electron configuration $j$ whose 
corresponding CI coefficient $|c^{(\alpha)}_j| < 0.0005$ was excluded in the 
expansion of that particular pseudostate for the close-coupling calculations. 
The value of this cutoff was determined by starting from 0.01 and decreasing it 
incrementally until negligible differences in the results were observed. This 
is contrast to the CI coefficient cutoff value of 0.02 used by 
\citet{Wang_2013_87}.

For a specified incident energy, the CC equations in Eq. (\ref{cceqs}) are 
solved for each impact parameter along a discretised $z$-grid using the 
Runge-Kutta method, where $z=vt$ is the position of the antiproton along the 
$z$-axis at time $t$. The extent of this grid is defined as $[-z_{\rm max}, 
z_{\rm max}]$. The number of points that discretise this grid and its extent 
are carefully chosen to ensure that the resulting cross sections are stable 
with respect to varying these parameters to three significant figures. In these 
calculations we find that setting $z_{\rm max} = 100$ a.u. and discretising the 
$z$-grid into 600 exponentially distributed points (denser set of points about 
the $z=0$ one) was sufficient. A vital check of the quality of our calculations 
is the norm of the scattering wavefunction. We ensure that, at all values of 
$z$ apart of the grid, $|\Psi^{+}_{i}|^2=1$ to within three significant 
figures. The number of impact parameters used as well as the largest impact 
parameter, $b_{\rm max}$, varies greatly depending on the incident energy. At 
10 keV, 66 impact parameter points with $b_{\rm max} = 15$ a.u., were required 
to ensure the weighted probabilities in the integrand of Eq. 
(\ref{cross_section}) vary smoothly and falloff a least three orders of 
magnitude from the maximum value. At 1000 keV, however, 94 impact parameter 
points with $b_{\rm max} = 50$ a.u. were necessary to achieve the same outcome.

\section{Results and discussion} \label{sec:results}
\begin{figure}[t]
    \centering
    \includegraphics{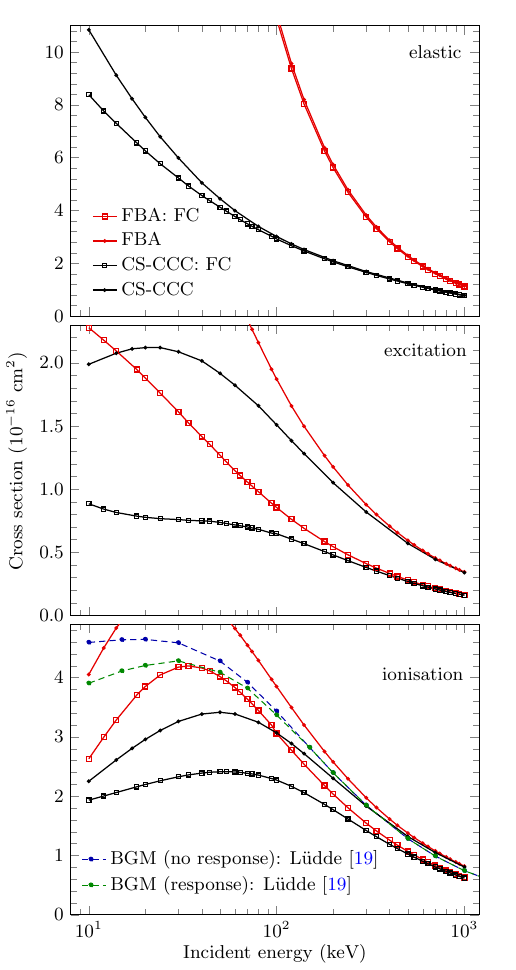}
    \caption{Cross sections for elastic scattering (top panel),
        total excitation (middle panel), and ionisation (bottom panel) in
        $\bar{p} + \ce{C}$ collisions. The present CS-CCC and FBA results 
        obtained using both the FC and MC target structure models are shown. 
        The BGM results by \citet{Ludde_2021_104} for net ionisation are
        also shown for comparison.
    }
    \label{fig1}
\end{figure}

Figure \ref{fig1} presents cross sections for elastic-scattering (top panel), 
total excitation (middle panel) and ionisation (bottom panel) as 
functions of the incident projectile energy. The total excitation cross 
section is defined as the sum of the cross sections for excitation of all the 
states below the first ionisation threshold. To better understand the role the 
accuracy of the target structure model plays in the scattering cross sections, 
we present CS-CCC results obtained using both the FC and MC models. 
Furthermore, for comparison, we also present both FC and MC results obtained 
using the first Born approximation (FBA). Two previous calculations for the net 
ionisation cross section using the BGM method with the no-response and response 
approaches \cite{Ludde_2021_104} are also shown.

We note that the CS-CCC results (both the FC and MC ones) merge with the 
corresponding FBA ones at the highest incident energies. However, it is clear 
that the CS-CCC and FBA results for the elastic-scattering cross section do not 
completely merge until beyond 1000 keV. We see that the CS-CCC excitation and 
ionisation cross sections merge with the corresponding FBA ones beyond about 
300 keV. 

Comparing the FC and MC cross sections resulting from the CS-CCC calculations 
with the corresponding FBA ones, we see minor differences in the 
elastic-scattering cross sections above 100 keV. Below this incident energy, 
however, the FC and MC results even for elastic scattering begin to deviate 
significantly, resulting in a 20\% difference between the two at 10 keV. 
Unlike the elastic-scattering cross section, the FC and MC results for 
excitation and ionisation differ significantly across the entire incident 
energy range considered. More specifically, the MC results sit above the FC 
ones at all energies. At 30 keV, the MC CS-CCC results are 2.75 times larger 
than the FC ones for total excitation. Similarly, at 50 keV, the MC ionisation 
cross section is 1.4 times larger than the FC one. Such significant differences 
in the cross sections for excitation and ionisation resulting from the FC and 
MC approaches can be understood by comparing the dipole polarisabilities from 
the two models. In the FC model we obtain a value of 6.98 $a_{0}^{3}$, while as 
previously mentioned in Sec. \ref{sec:details}, the MC model produces a value 
of 11.66 $a_{0}^{3}$ which is accurate and 1.67 times larger than the FC one. 
Putting all of these observations together, it is clear that modelling 
$\bar{p} + \ce{C}$ collisions using the FC target-structure model is not 
sufficient to accurately produce the elastic-scattering, excitation, and 
ionisation cross sections. There are two main reasons for this. The first is 
that the FC approach does not account for all of the valence-valence and 
core-valence electron correlations, which are crucial and must be taken into 
account when modelling the collisions at lower incident energies. The second 
reason is that the FC method does not include core-excitation and autoionising 
states in the close-coupling calculations. As we will see below, they make up a 
significant portion of the total excitation and ionisation cross sections.

As can be seen in the bottom panel of Fig. \ref{fig1}, the MC CS-CCC results  
for the ionisation cross section are lower than both sets of BGM calculations 
by \citet{Ludde_2021_104} below 300 keV. The largest difference between the MC 
CS-CCC results and the BGM ones with the no-response model occurs at 10 keV,
where the former is about 50\% smaller than the latter. There is a smaller, 
albeit still significant, difference between the MC CS-CCC and BGM results with 
the response model, which is at most about 40\%. As the incident energy 
increases above 300 keV, the difference between the three sets of results 
decreases. We note that the BGM calculations were done using an 
independent-particle model as well as the independent-event model to calculate 
the net ionisation \cite{Ludde_2021_104}.

\begin{figure}[t]
    \centering
    \includegraphics{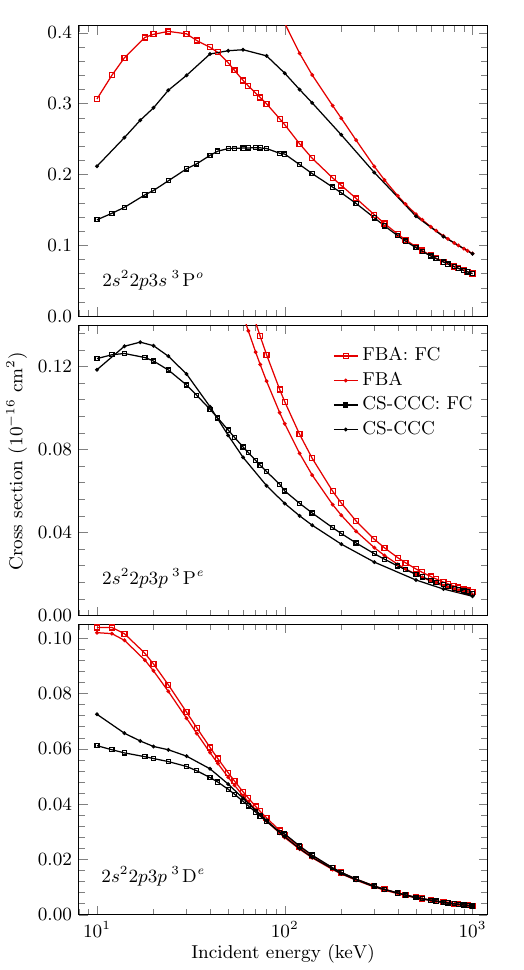}
    \caption{Cross sections of excitation to the \ce{$2s^22p3s$ ^3P^{o}} 
        (top panel), \ce{$2s^22p3p$ ^3P^{e}} (middle panel), and 
        \ce{$2s^22p3p$ ^3D^{e}} (bottom panel) states of C. The present CS-CCC 
        and FBA results obtained using both the FC and MC target structure 
        models are shown.
    }
    \label{fig2}
\end{figure}

The cross sections for target-excitation to states of C which can be described 
using both the FC and MC approaches are presented in Fig. \ref{fig2}
(obviously, the MC spectrum is much richer). As done in Fig. \ref{fig1}, for 
comparison FBA calculations with both target structure models are also
presented. The top panel shows the cross sections for excitation of the 
\ce{$2s^22p3s$ ^3P^{o}} state, the middle panel for the 
\ce{$2s^22p3p$ ^3P^{e}} state, and the bottom panel for the
\ce{$2s^22p3p$ ^3D^{e}} state. Similar to what we have already seen in Fig. 
\ref{fig1}, both the FC and MC sets of results from the FBA calculations merge 
with the CS-CCC ones at high incident energies for all the transitions. 
Furthermore, as can be seen in the bottom panel, the CS-CCC and FBA cross 
sections (both FC and MC) for excitation of the \ce{$2s^22p3p$ ^3D^{e}} state 
all merge together at about 70 keV incident energy and above. Similarly, we see 
the same behaviour for excitation of the \ce{$2s^22p3p$ ^3P^{e}} state, 
although at a much higher incident energy, just below 1000 keV. At the same 
time, looking at the high energy limit for the cross section corresponding to 
the excitation of the \ce{$2s^22p3s$ ^3P^{o}} state, we see that the FC and MC 
approaches do not merge at all. As this transition is a dipole-allowed one, we 
can look at the oscillator strength for this transition to understand why this 
is the case. The FC model produces an oscillator strength for the 
\ce{$2s^22p^2$ ^3P^{e}} $\to$ \ce{$2s^22p3s$ ^3P^{o}} transition of 0.102, 
meanwhile as seen in Table \ref{tab:oscillator_strengths}, the MC model 
produces a value of 0.130, which is in a much better agreement the accepted 
value \cite{NIST_ASD}. It is well known that in the high energy limit, the 
FBA cross section for dipole allowed transitions is proportional to the 
corresponding oscillator strength \cite{Bransden_1992_00}. Therefore, we expect 
to see roughly a similar level of difference between the cross sections using 
the FC and MC approaches as those seen in their respective oscillator 
strengths. This is indeed found to be the case since the oscillator strength 
from the MC model is 1.27 times larger than the FC one, meanwhile the 
corresponding cross section is 1.47 times larger. 

\begin{figure}[t]
    \centering
    \includegraphics{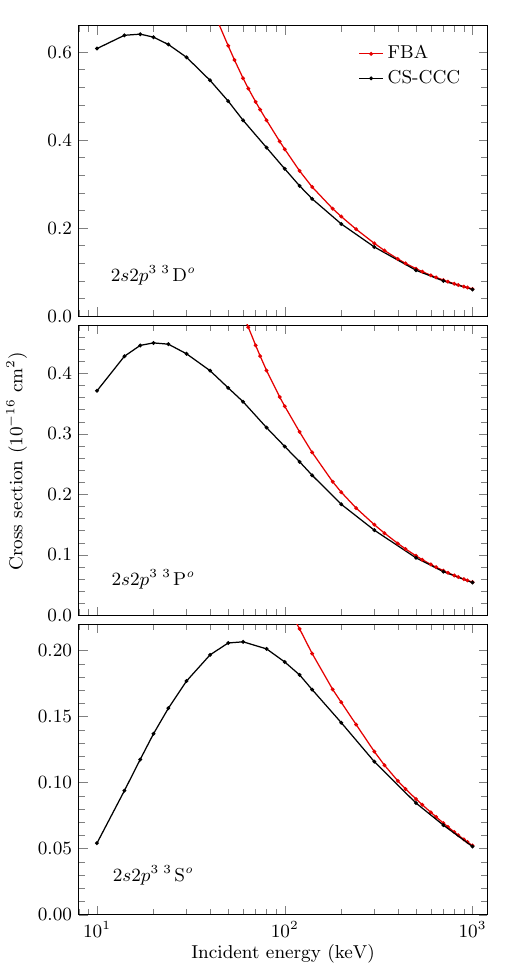}
    \caption{Cross sections of excitation to the \ce{$2s2p^3$ ^3D^{o}} 
        (top panel), \ce{$2s2p^3$ ^3P^{o}} (middle panel), and 
        autoionising \ce{$2s2p^3$ ^3S^{o}} (bottom panel) states of C. The present CS-CCC 
        and FBA results obtained using both the FC and MC target structure 
        models are shown.
    }
    \label{fig3}
\end{figure}

In Fig. \ref{fig3} we present the cross sections of excitation to the 
\ce{$2s2p^3$ ^3D^{o}}, \ce{$2s2p^3$ ^3P^{o}}, and \ce{$2s2p^3$ ^3S^{o}}
states of the C target. These collectively represent a set of core-excitation
states, where one of the inner-shell electron is excited to the valence shell.
Alongside the CS-CCC results, we also give the present FBA calculations. 
Here, there are no FC results to compare to, as the FC model does not include 
core-excited states. For these transitions, the CS-CCC and FBA results also 
merge at high incident energies like observed in the previous figures,
displaying a consistent behaviour in all the results we have presented.

An interesting observation can be made by comparing the magnitude of the cross 
sections for excitation to the \ce{$2s2p^3$ ^3D^{o}} and \ce{$2s2p^3$ ^3P^{o}} 
states to the total excitation cross section, shown in the middle panel of
Fig. \ref{fig1}. For example, at 20 keV incident energy, the sum of 
these two core-excitation cross sections is $1.08\times10^{-16}$ cm$^{2}$,
whereas the total excitation cross section is $2.07\times10^{-16}$ cm$^{2}$. 
This shows that at the peak of this cross section, the contribution from 
excitation of core electrons can be as high as 50\%. This seems to indicate 
that at least at lower incident energies, one of the main reasons why the FC 
and MC results, shown in the middle panel of Fig. \ref{fig1}, differ is the 
fact that the FC model does not include these important states in the 
coupled-channel calculations. Lastly, it is worth pointing out that the 
\ce{$2s2p^3$ ^3S^{o}} state is autoionising, meaning that it is situated above 
the first ionisation threshold of the C atom and contributes to the ionisation 
cross section presented in the bottom panel of Fig. \ref{fig1}. Thus, this 
excitation to this state alone is responsible for about 7\% of the ionisation 
cross section at its peak, further emphasising the importance of having a MC 
description of the C atom when modelling $\bar{p} + \ce{C}$ collisions.

\section{Conclusions and outlook} \label{sec:conclusions}
In this work we have developed a single-centre Coulomb-Sturmian-based 
convergent close-coupling method to study collisions with arbitrary 
multi-electron targets. We have done this using the $LS$ coupling scheme and by 
introducing a new atomic structure code that seamlessly incorporates into our 
semi-classical CS-CCC code suite. This software is capable of generating 
structure models with an arbitrary number of frozen and active electrons making 
it very robust and flexible. Using the set of pseudostates generated by this 
program, we can then expand the total scattering wavefunction that describes a 
heavy ion collision with the target atom of interest. Substituting this 
expansion into the full Hamiltonian of the collision system, we obtain a set of 
coupled-channel equations for the expansion coefficients which is solved
to determine the cross sections for any collision process of interest.

To demonstrate the capabilities of this new extension to the CS-CCC method, we
have applied it to study $\bar{p} + \ce{C}$ collisions. The structure model we 
developed to represent the C atom agree very well with the NIST data available 
\cite{NIST_ASD} for both the excitation energies and oscillator strengths of 
the first several excited states. Furthermore, the dipole polarisability of the 
ground state of the C atom produced by this model is within 3\% of the accepted 
value. The practical advantage of our structure model is that, in addition to 
accurate bound states, it generates pseudostates which span the continuous 
spectrum indicating that the present structure model is accurate and suitable
for subsequent close-coupling calculations to be performed. Within the incident 
energy range of 10 keV to 1000 keV, we have presented cross sections for 
elastic scattering, total excitation, and ionisation. We have also 
presented a set of state-resolved cross sections for excitation to states that 
can be described using both the FC and MC models and some core-excitation 
states which can only be described using the MC model. 

Carbon is one of the most abundant elements in the universe. It is also one of 
the common atomic building blocks in biologically relevant molecules. This 
makes studying collisions involving C, and molecular targets that contain 
carbon atoms, crucial for the further development of hadron therapy for cancer 
treatment \cite{Erdmann_2017_19}. The present results can be used to benchmark 
and encourage further studies of $\bar{p} + \ce{C}$ collisions. This newly 
developed theory is the first necessary step towards studying collisions 
involving other projectiles such as protons and other positively charged ions, 
where electron capture by the projectile is possible. Furthermore, this 
development of the CS-CCC method opens the door to studying a whole new set of 
ion collisions with other multi-electron targets, including impurity ions, of 
interest to plasma diagnostics \cite{Snipes_2024_41}, astrophysics 
\cite{Thompson_1996_29} and hadron therapy applications \cite{Nikjoo_2008_10} 
which was not possible with the previous implementations of the CCC approach to 
ion-atom scattering. Finally, it must be emphasised that even the current 
single-centre version of the CS-CCC method can be used to calculate electronic 
excitations and electron loss in proton and other highly charged ion collisions 
with multi-electron targets by simply increasing the basis size. When the size 
of the basis is large enough, the cross sections for excitation of the lowest 
states and total electron loss will be sufficiently accurate.

\clearpage

\begin{acknowledgments}
    This work was supported by the Australian Research Council. We acknowledge 
    the resources provided by Pawsey 
    Supercomputer centre, and the National Computing Infrastructure. N.W.A. 
    acknowledges support through Australian Government Research Training Program 
    Scholarships. We thank Prof. Tom Kirchner for providing the theoretical data 
    of Ref. \cite{Ludde_2021_104} in tabulated form.
\end{acknowledgments}

%

\end{document}